\documentclass[sigconf,anonymous=false]{acmart}
\AtBeginDocument{%
  }

\usepackage{booktabs}
\usepackage{multirow}

\begin{document}

\title{Beyond Static Collision Handling: Adaptive Semantic ID Learning for Multimodal Recommendation at Industrial Scale}

\author{Yongsen Pan}
\authornote{Both authors contributed equally to this research.}
\affiliation{%
  \institution{University of Electronic Science and Technology of China}
  \city{Chengdu}
  \country{China}
}

\author{Yuxin Chen}
\authornotemark[1]
\affiliation{%
  \institution{Kuaishou Technology}
  \city{Beijing}
  \country{China}
}
\email{chenyuxin06@kuaishou.com}

\author{Zheng Hu}
\affiliation{%
  \institution{University of Electronic Science and Technology of China}
  \city{Chengdu}
  \country{China}
}

\author{Xu Yuan}
\author{Daoyuan Wang}
\affiliation{%
  \institution{Kuaishou Technology}
  \city{Beijing}
  \country{China}
}

\author{Yuting Yin}
\author{Songhao Ni}
\affiliation{%
  \institution{Kuaishou Technology}
  \city{Beijing}
  \country{China}
}

\author{Hongyang Wang}
\affiliation{%
  \institution{Kuaishou Technology}
  \city{Beijing}
  \country{China}
}

\author{Jun Wang}
\affiliation{%
  \institution{Kuaishou Technology}
  \city{Beijing}
  \country{China}
}

\author{Fuji Ren}
\authornote{Corresponding author.}
\affiliation{%
  \institution{University of Electronic Science and Technology of China}
  \city{Chengdu}
  \country{China}
}
\email{renfuji@uestc.edu.cn}

\author{Wenwu Ou}
\affiliation{%
  \institution{Kuaishou Technology}
  \city{Beijing}
  \country{China}
}

\begin{abstract}
Modern recommendation systems operate over massive catalogs of multimodal items whose semantics strongly affect downstream recommendation quality, making scalable item identification a representation problem that must balance compactness for efficient recommendation with semantic fidelity. Semantic IDs (SIDs) offer a practical solution to this representation challenge by mapping multimodal item signals into short discrete token sequences, thereby providing a compact yet semantically informed item interface for retrieval, ranking, and emerging generative recommendation. However, learning effective SIDs remains challenging. One key challenge is collision: different items may be assigned identical or highly confusable codes in the discrete space. Existing methods mainly improve SID learning through better quantization or fixed overlap regularization, but still lack an adaptive treatment of discrete ambiguity for recommendation-oriented SID learning. To address this issue, we propose AdaSID, an adaptive semantic ID learning framework for recommendation. AdaSID regulates overlaps through a two-stage adaptive process. In the first stage, it adaptively relaxes repulsion for each observed overlap based on semantic compatibility, determining whether it should continue to receive repulsion or be preserved as semantically admissible sharing. In the second stage, it adaptively allocates the remaining regulation pressure according to local collision load and training progress, so that SID learning remains aligned with both discrete-space structure and recommendation-oriented optimization. Extensive offline and online experiments validate AdaSID. On two public benchmarks, AdaSID improves Recall and NDCG by around \textbf{4.5\%} on average over strong baselines while also improving codebook utilization and SID diversity. In industrial validation on Kuaishou e-commerce, AdaSID further demonstrates practical value. An online A/B test on the short-video retrieval model covering tens of millions of users yields statistically significant gains in key business metrics, including a \textbf{0.98\%} improvement in GMV, while industrial ranking evaluation also shows consistent AUC improvements.
\end{abstract}




\maketitle

\section{Introduction}

Modern recommendation systems increasingly operate on items equipped with rich content signals such as text, images, and videos. Effective item representation is therefore a fundamental requirement in both generative and discriminative recommendation. Traditional sparse IDs are efficient, but they carry little multimodal semantic information and treat items as isolated symbols \cite{10.1145/3746027.3754937}, which limits generalization across related items \cite{10.1145/3640457.3688190}. This limitation becomes more pronounced in dynamic recommendation scenarios such as short-video platforms, where items are continuously created, consumed, and replaced. In such settings, sparse-ID-based embeddings are prone to representation drift over time and are especially difficult to keep well trained for tail and newly arrived items \cite{10.1145/3705328.3748123}. Direct multimodal item representations are more expressive, yet they do not serve as stable item identifiers for recommendation and are not directly optimized into stable item identifiers under downstream recommendation supervision \cite{10.1145/3746252.3761502}. To bridge this gap, Semantic IDs (SIDs) have recently attracted growing attention in both academia and industry \cite{10.1145/3746252.3761439,10.1145/3746252.3761529,10.1145/3664647.3680574}. By converting multimodal item features into short discrete token sequences, SIDs provide a practical balance between semantic richness, compactness, and learnability for modern recommendation pipelines \cite{10.5555/3692070.3692964}. 

Despite these advantages, learning effective SIDs remains challenging. A central difficulty is collision: during the projection from multimodal item features to discrete SIDs, different items may be assigned identical or highly confusable SIDs in the discrete space \cite{10.1145/3627673.3679569,10.1145/3726302.3729989}. This issue is especially consequential in recommendation, because SIDs serve not only as compressed representations, but also as the discrete interface through which downstream models differentiate items. When semantically or behaviorally mismatched items share similar SIDs, the downstream model is forced to update the same discrete representations according to inconsistent supervision signals, which can lead to conflicting optimization and degraded recommendation quality \cite{li2025survey}. As a result, collision can directly undermine the recommendation utility of the learned SID space.

\begin{figure}[t]
    \centering
    \includegraphics[width=0.85\linewidth]{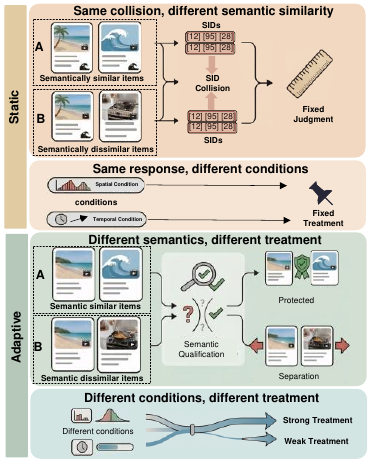}
    \caption{Motivation of AdaSID. Static overlap regulation uses fixed judgment and fixed treatment, whereas AdaSID instead performs adaptive semantic qualification to determine whether an observed overlap should continue to receive repulsion, and then allocates the remaining regulation pressure adaptively.}
    \label{fig:illustration}
\end{figure}

Existing studies have improved SID learning mainly by enhancing the quality of discrete representations. Some focus on improving quantization quality, codebook utilization, or training stability, thereby mitigating collisions indirectly through better-structured discretization \cite{van2017neural,DBLP:conf/cvpr/LeeKKCH22,Mentzer2024FiniteSQ}. Others introduce explicit overlap-aware objectives to suppress harmful collisions in the SID space \cite{hu2026stop,DBLP:journals/corr/abs-2508-04618}. These advances are important, but their treatment of overlaps remains largely static once collisions are observed. The weakness of such static treatment is twofold. First, overlaps are judged under a fixed standard, without adequately distinguishing harmful ambiguity from semantically compatible sharing in the multimodal space. Second, once an overlap is treated as problematic, it is usually regulated with a largely fixed mode of response. As a result, overlap regulation can fail to distinguish what should be separated from what can be shared, and can further remain insensitive to the varying conditions under which retained overlaps should be handled, thereby weakening the recommendation utility of the learned SID space, as illustrated in Figure~\ref{fig:illustration}.

To address these limitations, we propose AdaSID, an adaptive SID learning framework for recommendation. AdaSID casts overlap regulation as a two-stage adaptive process. \textbf{The first stage} determines from the semantic perspective whether an observed overlap should continue to receive repulsion or can be preserved as semantically admissible sharing. \textbf{The second stage} then applies adaptive pressure allocation to the remaining overlaps, both spatially and temporally. Spatially, it regulates how separation pressure is distributed in the SID space through load-adaptive collision strengthening. Temporally, it adjusts the optimization emphasis of this regulation over training through progress-adaptive objective rebalancing. In this way, AdaSID turns overlap handling from a fixed response into recommendation-oriented adaptive regulation. Extensive offline and online experiments validate the proposed framework. On two public benchmarks, AdaSID improves Recall and NDCG by around 4.5\% on average over strong baselines while also improving SID codebook quality. In industrial validation on Kuaishou e-commerce, AdaSID further yields consistent gains in both online retrieval A/B tests and offline ranking, demonstrating measurable value in real recommendation traffic. The main contributions of this work are as follows:
\begin{itemize}
    \item We address static overlap handling in existing SID learning. To this end, we propose AdaSID, an adaptive framework that casts overlap regulation as a two-stage process. In the first stage, AdaSID determines, based on semantic consistency, whether an observed overlap should continue to receive repulsion, thereby distinguishing harmful ambiguity from semantically admissible sharing during SID learning.

    \item Building on this semantic qualification step, we develop the second stage of AdaSID as an adaptive pressure allocation design for the retained overlaps. This design regulates overlap handling both spatially and temporally, through load-adaptive collision strengthening and progress-adaptive objective rebalancing, respectively.

    \item We conduct comprehensive offline and industrial evaluations of AdaSID, including recommendation accuracy, SID-space diagnostics, online retrieval A/B tests, and offline ranking studies. The results consistently show that AdaSID improves both recommendation performance and the quality of the learned SID space in real recommendation scenarios.
\end{itemize}

\section{Related Work}
\subsection{Multimodal Item Representation for Recommendation}

Learning item representations from multimodal content has become an important direction in recommendation systems. Modern items are often associated with heterogeneous signals such as text, images, and videos \cite{10.1145/3637528.3671473,10.1145/3746027.3755540,10.1145/3746027.3755781,10.1145/3637528.3672041}. A large body of work has shown that leveraging these signals can enrich item semantics beyond sparse IDs and improve recommendation quality, especially in cold-start and content-sensitive scenarios \cite{10.5555/3015812.3015834,10.1145/3343031.3351034,10.1145/3640457.3688098}. Representative techniques include cross-modal fusion \cite{10.1145/3474085.3475259}, contrastive alignment \cite{10.1145/3512527.3531378}, pretrained encoders \cite{10.1145/3682075}, and modality-aware interaction modeling \cite{10.1145/3543507.3583251}. These studies have substantially improved multimodal item understanding and representation learning for recommendation. However, most of them still model items as continuous embeddings, whereas SID learning focuses on constructing compact discrete item identifiers that can better support discrete indexing, token-based generation, and reusable item interfaces across recommendation stages \cite{rajput2023recommender,10.1145/3640457.3688190}. This distinction has motivated a separate line of research on semantic ID learning.

\subsection{SID Learning for Recommendation}

SID learning has recently become an important direction in recommendation. It maps item semantics into short discrete token sequences and provides a unified item interface for retrieval, ranking, and emerging generative recommendation models \cite{li2025survey}. Existing studies have explored this paradigm in both discriminative and generative settings, showing that SIDs can better balance semantic richness, vocabulary efficiency, and deployment flexibility than conventional sparse IDs or purely textual item representations \cite{10.1145/3543507.3583434,10.1145/3640457.3688190,10597986}. Methodologically, prior work mainly differs in how discrete codes are constructed. Representative routes include product quantization \cite{10.1145/3543507.3583434}, residual quantization \cite{10.1145/3746252.3761502,chen2025onesearch}, and hierarchical or tree-structured tokenization \cite{10.1145/3701716.3715234,10.1145/3711896.3736979}. More recent studies further incorporate collaborative signals \cite{10597986}, multimodal alignment \cite{10.1145/3711896.3737275}, or end-to-end downstream supervision \cite{10.1145/3746252.3761529,10.1145/3726302.3729989} so that learned IDs better reflect recommendation relevance rather than content similarity alone. Despite these advances, existing methods mainly improve SID construction through better discretization, utilization, or downstream alignment, while overlap handling is usually treated in a fixed or secondary manner once collisions arise \cite{DBLP:journals/corr/abs-2508-04618,hu2026stop}. As a result, adaptive overlap regulation for recommendation-oriented SID learning remains underexplored.

\section{Notation and Problem Definition}

\subsection{Notation}
\label{sec:notion}
Let $\mathcal{D}=\{(x_i^{\mathrm{tr}}, x_i^{\mathrm{ta}})\}_{i=1}^{N}$ denote the training set of collaborative item pairs, constructed from collaborative signals such as Swing \cite{DBLP:journals/corr/abs-2010-05525} in this work. Here, $x_i^{\mathrm{tr}}$ and $x_i^{\mathrm{ta}}$ denote the multimodal inputs of the trigger item and the target item in the $i$-th pair, respectively. Each item input $x$ denotes a multimodal item feature extracted from heterogeneous content signals such as text, image, and video by a pretrained multimodal model. On top of these extracted item features, a shared encoder $f_{\theta}(\cdot)$ maps $x_i$ to a continuous representation $z_i=f_{\theta}(x_i)\in\mathbb{R}^{d}$, where $d$ is the embedding dimension and $\theta$ denotes the encoder parameters. We adopt an $L$-layer residual vector quantizer \cite{DBLP:conf/cvpr/LeeKKCH22} with codebooks $\{\mathcal{C}^{(l)}\}_{l=1}^{L}$, where $\mathcal{C}^{(l)}=\{c_{1}^{(l)},\dots,c_{K}^{(l)}\}$ and each codeword $c_{k}^{(l)}\in\mathbb{R}^{d}$. Here, $K$ denotes the codebook size of each layer. For item $i$, the residual representation is initialized as $r_i^{(0)}=z_i$. At layer $l$, the quantizer selects the nearest codeword index $s_i^{(l)}\in\{1,\dots,K\}$, obtains the corresponding quantized vector $q_i^{(l)}=c_{s_i^{(l)}}^{(l)}$, and updates the residual as $r_i^{(l)}=r_i^{(l-1)}-q_i^{(l)}$. After $L$ layers, the quantized embedding is $\hat{z}_i=\sum_{l=1}^{L} q_i^{(l)}$, and the resulting SID is the discrete index sequence $s_i=[s_i^{(1)},s_i^{(2)},\ldots,s_i^{(L)}]$. A decoder $h_{\phi}(\cdot)$ reconstructs the original item representation from the quantized embedding, yielding $\tilde{x}_i=h_{\phi}(\hat{z}_i)$, where $\phi$ denotes the decoder parameters. Since AdaSID is learned from item pairs, we further define several pairwise quantities used in the method section. For two items $i$ and $j$, their SID overlap depth is defined as $o_{ij}=\sum_{l=1}^{L}\mathbb{I}[s_i^{(l)}=s_j^{(l)}]$, where $\mathbb{I}[\cdot]$ is the indicator function. A larger $o_{ij}$ indicates that the two items are closer in the discrete SID space. We use $\mathrm{sim}_{ij}$ to denote their semantic similarity in the continuous space, computed from encoder-side representations. Finally, $\tau\in[0,1]$ denotes the normalized training progress for progress-adaptive objective rebalancing.

\subsection{Problem Definition}

Given the collaborative pair set $\mathcal{D}$, our goal is to learn a SID mapping that converts each multimodal item input $x$ into a short discrete index sequence $s=[s^{(1)},\dots,s^{(L)}]$. The learned SIDs should satisfy two requirements simultaneously: they should preserve useful multimodal semantics under discretization, and they should remain well aligned with collaborative relations so as to support downstream recommendation. The main difficulty is that discrete overlaps are heterogeneous: the same overlap depth may play different roles in the learned SID space and therefore require different treatment during training. Accordingly, our objective is to learn a recommendation-oriented SID space in which discrete overlaps are regulated adaptively, rather than being handled uniformly once they are observed.

\section{Method}
In this section, we present AdaSID as a recommendation-oriented SID learning framework with two-stage adaptive overlap regulation. \textbf{The first stage}, semantics-adaptive overlap relaxation, determines whether an observed overlap should continue to receive repulsion, and \textbf{the second stage}, adaptive pressure allocation, regulates the retained overlaps across discrete regions and training progress through load-adaptive collision strengthening and progress-adaptive objective rebalancing.
\begin{figure*}[t]
    \centering
    \includegraphics[width=0.75\textwidth]{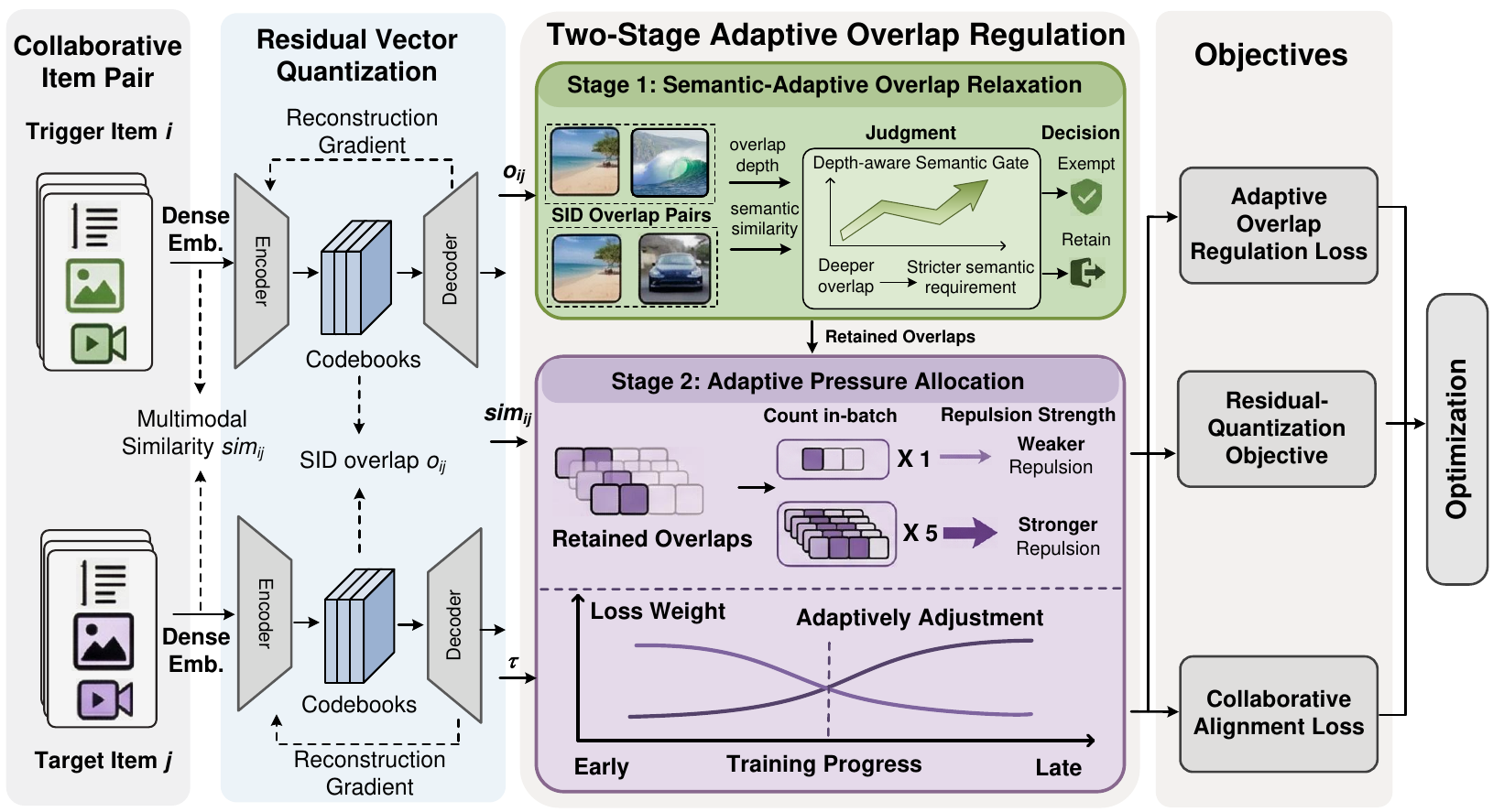}
    \caption{Overall framework of AdaSID. AdaSID maps collaborative trigger--target item pairs into quantized representations and Semantic IDs, and then optimizes them through a two-stage adaptive overlap regulation process: semantic-adaptive overlap relaxation first determines whether repulsion should be relaxed, after which load-adaptive collision strengthening and progress-adaptive objective rebalancing regulate the remaining overlaps across discrete regions and training time.}
    \label{fig:framework}
\end{figure*}

\subsection{Overall Framework}

AdaSID learns SIDs from collaborative item pairs by coupling multimodal discretization with recommendation-oriented alignment. Given a training pair $(x_i^{\mathrm{tr}}, x_i^{\mathrm{ta}})$, the trigger item and the target item are first projected by a shared encoder to continuous representations, then converted by the residual quantizer into discrete SIDs, and finally reconstructed by a decoder from the quantized embeddings. Let $Q(\cdot)$ denote the $L$-layer residual quantization process defined in Section~\ref{sec:notion}. The forward process is written as:
\begin{equation}
\begin{aligned}
z_i^{\mathrm{tr}} &= f_{\theta}(x_i^{\mathrm{tr}}), \qquad (\hat{z}_i^{\mathrm{tr}}, s_i^{\mathrm{tr}})=Q(z_i^{\mathrm{tr}}), \qquad \tilde{x}_i^{\mathrm{tr}} = h_{\phi}(\hat{z}_i^{\mathrm{tr}}), \\
z_i^{\mathrm{ta}} &= f_{\theta}(x_i^{\mathrm{ta}}), \qquad (\hat{z}_i^{\mathrm{ta}}, s_i^{\mathrm{ta}})=Q(z_i^{\mathrm{ta}}), \qquad \tilde{x}_i^{\mathrm{ta}} = h_{\phi}(\hat{z}_i^{\mathrm{ta}}).
\end{aligned}
\end{equation}
Here, $\hat{z}_i^{\mathrm{tr}}$ and $\hat{z}_i^{\mathrm{ta}}$ are the quantized embeddings, while $s_i^{\mathrm{tr}}$ and $s_i^{\mathrm{ta}}$ are the corresponding SIDs. Since nearest-codeword lookup is non-differentiable, we adopt the straight-through estimator (STE) \cite{bengio2013estimating} to pass gradients from the quantized embeddings to the encoder-side representations during back-propagation.

The forward pipeline above defines the tokenizer backbone. AdaSID introduces adaptive overlap regulation over in-batch SID overlaps so that the learned discrete space is regulated through adaptive mechanisms rather than by a fixed rule. To formalize this process, given a mini-batch, we first collect the set of candidate overlap pairs:
\begin{equation}
\mathcal{P}=\left\{(i,j)\mid i\neq j,\; o_{ij}>0\right\}.
\end{equation}
Here, $o_{ij}$ is the SID overlap depth defined in Section~\ref{sec:notion}. Following prior overlap-aware SID regularization in SID learning \cite{hu2026stop}, we instantiate collision supervision as a margin-based repulsion on encoder-side representations. Let
\begin{equation}
d_{ij}^{\mathrm{c}}=1-\frac{z_i^{\top}z_j}{\|z_i\|_2\|z_j\|_2}
\end{equation}
denote the cosine distance between items $i$ and $j$. A generic base collision penalty is then written as:
\begin{equation}
\ell_{ij}^{\mathrm{col}}=\max\!\left(0,\,m_{ij}-d_{ij}^{\mathrm{c}}\right),
\label{eq:base_collision}
\end{equation}
where $m_{ij}$ is a margin determined by the overlap type or severity of pair $(i,j)$. This base penalty serves as the common collision form used throughout AdaSID. 

\subsection{Stage One: Semantic-Adaptive Overlap Relaxation}

The first stage of AdaSID determines, based on semantic consistency, whether an observed overlap should continue to receive repulsion. Collision regulation in recommendation-oriented SID learning should not only separate harmful ambiguity, but also preserve semantically admissible sharing. Some overlap pairs remain highly consistent in the continuous space and therefore need not be repelled as aggressively as clearly mismatched pairs. To this end, AdaSID adopts an adaptive overlap relaxation mechanism. More importantly, the relaxation criterion is not uniform: as discrete overlap becomes deeper, relaxing repulsion should require progressively stronger semantic evidence. This graded design avoids a one-size-fits-all relaxation rule and makes the first-stage decision adaptive to overlap severity.

For each candidate pair $(i,j)\in\mathcal{P}$, we compute the semantic similarity between their encoder-side representations as:
\begin{equation}
\mathrm{sim}_{ij}
=
\frac{z_i^{\top} z_j}{\|z_i\|_2 \|z_j\|_2}.
\end{equation}
To encode the graded relaxation criterion, we design a depth-aware threshold vector:
\begin{equation}
\boldsymbol{\eta}=[\eta_{1},\eta_{2},\ldots,\eta_{L}],
\qquad
\eta_{1}\leq\eta_{2}\leq\cdots\leq\eta_{L},
\end{equation}
where $\eta_{o_{ij}}$ denotes the semantic threshold associated with overlap depth $o_{ij}$. The relaxation gate is then instantiated as:
\begin{equation}
g_{ij}
=
\mathbb{I}\!\left[\mathrm{sim}_{ij} \geq \eta_{o_{ij}}\right].
\end{equation}
This design avoids a one-size-fits-all relaxation rule. Shallow overlaps are relaxed under relatively mild semantic requirements, while deeper overlaps are relaxed only when the corresponding items remain highly consistent in the continuous space. As a result, AdaSID preserves semantically admissible sharing without allowing strong discrete overlaps to remain insufficiently regulated merely because they are close in the continuous space. This stage therefore determines which observed overlaps should continue to receive repulsion in the subsequent pressure-allocation stage.

\subsection{Stage Two: Adaptive Pressure Allocation}

After the first-stage semantic-adaptive overlap relaxation, AdaSID regulates the retained overlaps through adaptive pressure allocation. The key idea is that once repulsion is retained, it should not be applied uniformly. Instead, the second stage allocates regulation pressure from two complementary aspects. Spatially, it determines where stronger separation should be imposed in the SID space. Temporally, it adjusts how strongly collision regulation should influence the overall optimization as training proceeds. The following two subsections realize this second stage through load-adaptive collision strengthening and progress-adaptive objective rebalancing, respectively.

\subsubsection{Load-Adaptive Collision Strengthening}

The spatial side of adaptive pressure allocation is realized through load-adaptive collision strengthening. Once repulsion is retained after the first-stage semantic-adaptive overlap relaxation, AdaSID does not regulate all such overlaps with uniform strength. Instead, it considers where stronger correction should be placed in the discrete SID space. Overlap depth $o_{ij}$ captures how close two items are in the discrete SID space, but it does not describe how densely the same overlap pattern is reused within the current mini-batch. For recommendation-oriented SID learning, this regional concentration is important: when many pairs repeatedly collapse into a similar discrete neighborhood, the corresponding SID region becomes overloaded and its discriminability is weakened. AdaSID therefore performs load-adaptive collision strengthening by complementing overlap depth with local collision-load estimation and assigning stronger regularization to more congested regions. For each candidate pair $(i,j)\in\mathcal{P}$, we define its collision signature as the layer-wise overlap pattern:
\begin{equation}
\kappa_{ij}
=
\left[
\mathbb{I}\!\left[s_i^{(1)}=s_j^{(1)}\right],
\mathbb{I}\!\left[s_i^{(2)}=s_j^{(2)}\right],
\dots,
\mathbb{I}\!\left[s_i^{(L)}=s_j^{(L)}\right]
\right].
\end{equation}
The local collision load of pair $(i,j)$ is then measured by the number of overlap pairs in the mini-batch that share the same signature:
\begin{equation}
c_{ij}
=
\sum_{(u,v)\in\mathcal{P}}
\mathbb{I}\!\left[\kappa_{uv}=\kappa_{ij}\right].
\end{equation}
A larger $c_{ij}$ indicates that the same discrete overlap pattern is reused more frequently and therefore deserves stronger separation. We instantiate the load-adaptive coefficient $a_{ij}$ by a bounded monotone scaling function:
\begin{equation}
a_{ij}
=
g(c_{ij};\,f_{\min},f_{\max},d_{\max},\alpha),
\end{equation}
where $g(\cdot)$ increases with the local collision load $c_{ij}$, $f_{\max}$ controls the maximum strengthening factor, $d_{\max}$ controls the effective load range before saturation, and $\alpha$ controls the growth sharpness. This design keeps the strengthening factor bounded while still assigning larger penalties to highly congested overlap patterns.

Load-adaptive strengthening thus equips collision regulation with regional context beyond pairwise overlap depth. While $o_{ij}$ describes the proximity of one pair in the SID space, $c_{ij}$ and $a_{ij}$ characterize how heavily the same overlap pattern accumulates in the current batch. Their combination enables AdaSID to focus stronger separation pressure on overloaded discrete neighborhoods, thereby improving the separability and utilization of the learned SID space.

\subsubsection{Progress-Adaptive Objective Rebalancing}

The temporal side of adaptive pressure allocation concerns how the optimization emphasis should evolve during training. Early in training, the encoder and codebooks are still forming a usable discrete structure, so stronger collision-related regulation helps prevent the emerging SID space from collapsing into crowded regions. As training proceeds and the discrete structure becomes more stable, recommendation-oriented alignment should gradually play a larger role. AdaSID captures this shift through progress-adaptive objective rebalancing, so that the optimization emphasis evolves with training progress rather than remaining fixed throughout training.

Let $t$ denote the current training step. We define the normalized schedule progress as:
\begin{equation}
\tau
=
\mathrm{clip}\!\left(
\frac{t-T_{\mathrm{start}}}{T_{\mathrm{end}}-T_{\mathrm{start}}},
\,0,\,1
\right),
\end{equation}
where $T_{\mathrm{start}}$ and $T_{\mathrm{end}}$ denote the start and end steps of adaptive rebalancing. Based on $\tau$, AdaSID uses progress-dependent objective weights to adjust the optimization emphasis over training. In particular, collision-related regulation is emphasized more strongly in earlier training and is gradually relaxed as training proceeds, while collaborative alignment becomes increasingly important. This design allows the model to first establish a stable discrete structure and then progressively place more emphasis on recommendation-oriented alignment. In the implementation used in this paper, we instantiate this idea by decaying the collision weight toward a non-zero floor while increasing the collaborative weight:
\begin{equation}
\lambda_{\mathrm{col}}(\tau)=1-(1-\lambda_{\mathrm{col}}^{\min})\tau,
\qquad
\lambda_{\mathrm{cf}}(\tau)=\lambda_{\mathrm{cf}}^{\max}\tau.
\label{eq:stage_weight}
\end{equation}
This schedule keeps collision regulation more prominent when the SID space is still unstable, and gradually increases the role of collaborative alignment as training proceeds. In this way, AdaSID adapts not only which overlaps are regulated and how strongly they are regulated, but also how their role changes over the course of training.

\subsection{Overall Objective and Training Procedure}

In our implementation, the stage-specific mechanisms introduced above are combined into the adaptive collision term:
\begin{equation}
L_{\mathrm{col}}^{\mathrm{ada}}
=
\sum_{(i,j)\in\mathcal{P}}
a_{ij}\left(1-g_{ij}\right)\ell_{ij}^{\mathrm{col}}.
\label{eq:adaptive_collision}
\end{equation}
Here, $g_{ij}$ relaxes repulsion under semantic compatibility, while $a_{ij}$ modulates the strength of repulsion according to local collision load. Using the progress-adaptive weights in Eq.~\ref{eq:stage_weight}, the overall training objective is:
\begin{equation}
L
=
L_{\mathrm{rec}}
+
L_{\mathrm{rq}}
+
\lambda_{\mathrm{col}}(\tau)L_{\mathrm{col}}^{\mathrm{ada}}
+
\lambda_{\mathrm{cf}}(\tau)L_{\mathrm{cf}},
\label{eq:overall_objective}
\end{equation}
where $L_{\mathrm{rec}}$ is the reconstruction loss between the input multimodal feature and the decoded feature, $L_{\mathrm{rq}}$ is the standard residual-quantization objective, and $L_{\mathrm{cf}}$ is the collaborative alignment objective between paired items. In our implementation, $L_{\mathrm{cf}}$ is instantiated as an InfoNCE-style contrastive loss \cite{DBLP:journals/corr/abs-1807-03748} over the quantized embeddings.

During training, each mini-batch is encoded and quantized with STE, after which AdaSID computes overlap patterns, estimates load-adaptive factors and semantic relaxation gates, and applies progress-adaptive weights to the collision and collaborative terms. All parameters are optimized jointly by back-propagation. At inference time, AdaSID uses the learned encoder and residual quantizer to generate SIDs as in a standard SID pipeline, so the adaptive mechanisms improve SID learning during training without introducing extra online complexity.

\section{Experiments}

\begin{table}[t]
\centering
\caption{Statistics of the offline datasets.}
\label{tab:dataset_stats}
\begin{tabular}{lcccc}
\toprule
Dataset & \#Users & \#Items & \#Interactions & Density \\
\midrule
Amazon-Toys   & 19,412 & 11,924 & 905,253   & 0.3911\% \\
Amazon-Beauty & 22,363 & 12,101 & 1,048,296 & 0.3874\% \\
\bottomrule
\end{tabular}
\end{table}

\begin{table*}[t]
\centering
\caption{Performance comparison of different SID tokenizers on the Toys and Beauty datasets. The best results are boldfaced, and the best baseline results are underlined.}
\label{tab:main_results}
\normalsize
\setlength{\tabcolsep}{4.5pt}
\renewcommand{\arraystretch}{1.0}
\begin{tabular}{lcccccccc}
\toprule
\multirow{2}{*}{Tokenizer} 
& \multicolumn{4}{c}{Toys} 
& \multicolumn{4}{c}{Beauty} \\
\cmidrule(lr){2-5} \cmidrule(lr){6-9}
& Recall@3 & NDCG@3 & Recall@5 & NDCG@5 
& Recall@3 & NDCG@3 & Recall@5 & NDCG@5 \\
\midrule
RQ-OPQ           & 0.0176 & 0.0152 & 0.0215 & 0.0178 & 0.0181 & 0.0152 & 0.0225 & 0.0170 \\
RQ-VAE           & 0.0164 & 0.0142 & 0.0197 & 0.0161 & 0.0161 & 0.0131 & 0.0206 & 0.0149 \\
Improved VQGAN   & 0.0191 & \underline{0.0164} & 0.0224 & 0.0177 & 0.0178 & 0.0146 & 0.0231 & 0.0167 \\
GRVQ             & 0.0170 & 0.0147 & 0.0192 & 0.0166 & 0.0189 & 0.0151 & 0.0246 & 0.0179 \\
Rotation Trick   & 0.0182 & 0.0157 & 0.0221 & 0.0183 & 0.0193 & \underline{0.0155} & 0.0245 & 0.0180 \\
SimRQ            & 0.0191 & 0.0160 & 0.0216 & 0.0175 & 0.0182 & 0.0147 & 0.0236 & 0.0169 \\
RQ-KMeans        & 0.0193 & 0.0160 & 0.0271 & 0.0187 & 0.0199 & 0.0151 & \underline{0.0271} & 0.0184 \\
QuaSID           & \underline{0.0195} & 0.0157 & \underline{0.0273} & \underline{0.0191} & \underline{0.0201} & \underline{0.0155} & 0.0268 & \underline{0.0186} \\
\midrule
AdaSID           & \textbf{0.0214} & \textbf{0.0175} & \textbf{0.0281} & \textbf{0.0202} & \textbf{0.0205} & \textbf{0.0164} & \textbf{0.0275} & \textbf{0.0190} \\
\bottomrule
\end{tabular}
\end{table*}

\subsection{Experimental Setup}

\subsubsection{Offline Datasets}

We conduct experiments on two public benchmarks from Amazon 2018 review datasets\footnote{https://nijianmo.github.io/amazon/}, namely \textit{Beauty} and \textit{Toys \& Games} (abbreviated as \textit{Toys}). Following prior work \cite{DBLP:conf/recsys/Geng0FGZ22}, we apply a standard 5-core filtering procedure, discarding users and items with fewer than five interactions. We then adopt a leave-one-out strategy to
split each dataset into training, validation, and test sets. Item textual fields include \textit{Title}, \textit{Brand}, \textit{Categories}, and \textit{Price}. We concatenate these fields and extract semantic item embeddings using Sentence-T5-XXL \cite{DBLP:conf/acl/NiACMHCY22}. The statistics of the offline experiments are shown in Table~\ref{tab:dataset_stats}.

\subsubsection{Evaluation Metrics}

\begin{figure*}[t]
    \centering
    \includegraphics[width=0.46\textwidth]{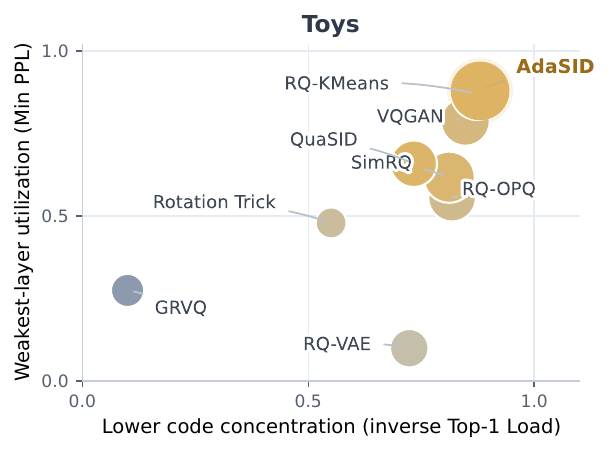}
    \includegraphics[width=0.46\textwidth]{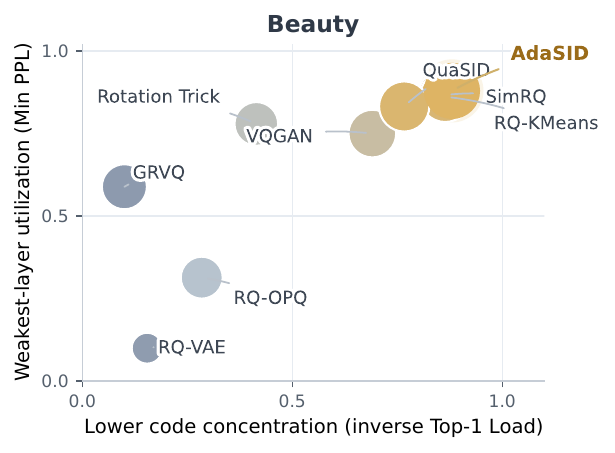}
    \caption{Discrete SID space landscape on two datasets. Each point denotes a tokenizer. The x-axis is inverse normalized Top-1 Load, and the y-axis is normalized Minimum Perplexity. Point size and color denote Mean Perplexity and SID Entropy. Points nearer the upper-right, and with larger and darker markers, indicate a more uniformly utilized and more diverse SID space.}
    \label{fig:sid_space_landscape}
\end{figure*}

For offline recommendation performance, we report Recall@$K$ and NDCG@$K$ with $K\in\{3,5\}$. Since all compared methods are evaluated under the same downstream TIGER backbone \cite{rajput2023recommender}, differences in these metrics can be attributed to the quality of the learned SIDs. To characterize the learned SID space, we further compute four codebook-side statistics on each dataset and visualize their joint behavior in Figure~\ref{fig:sid_space_landscape}. Specifically, \textbf{SID Entropy} measures the entropy of the full SID sequence distribution and reflects how broadly SID sequences occupy the discrete sequence space. \textbf{Average Perplexity} is the mean perplexity across codebook layers and reflects overall codebook utilization. \textbf{Minimum Perplexity} measures the weakest-utilized layer and serves as a stricter indicator of layer-wise under-utilization. \textbf{Mean Top-1 Load Ratio} measures the average fraction of samples assigned to the most frequently used code in each layer, where smaller values indicate weaker dominant-code concentration. In the visualization, these statistics are mapped to axis position, point size, and point color to reveal the overall structure of the learned SID space rather than any single metric in isolation. For industrial validation, we additionally report online business metrics and offline ranking AUC gain. Specifically, \textbf{Orders} measures completed transactions, \textbf{GMV} denotes gross merchandise volume, and \textbf{GPM} denotes gross merchandise volume per mille exposures. For the ranking module, we report \textbf{AUC gain} in percentage points (pp).

\subsubsection{Baselines}

We compare AdaSID against representative SID pre-training methods. To ensure fair comparison, all methods use the same downstream generative recommendation backbone and differ only in how SIDs are learned before downstream training. Our baselines fall into three categories. 
(1) \textbf{Standard quantization-based SID learning methods}, represented by vanilla RQ-VAE \cite{DBLP:conf/cvpr/LeeKKCH22}. 
(2) \textbf{Improved quantization or code-utilization methods}, including GRVQ \cite{DBLP:journals/corr/abs-2305-02765}, RQ-OPQ \cite{chen2025onesearch}, Rotation Trick \cite{DBLP:conf/iclr/FiftyJDILATR25}, Improved VQGAN \cite{DBLP:conf/iclr/YuLKZPQKXBW22}, SimRQ \cite{zhu2025addressing}, and RQ-KMeans \cite{10.1145/3746252.3761502}, which improve optimization stability, code usage, or quantization quality from different perspectives. 
(3) \textbf{Collision-aware SID learning methods}, represented by QuaSID \cite{hu2026stop}, which introduces explicit collision regularization during SID pre-training. 

\begin{table}[t]
\centering
\caption{Industrial validation on Kuaishou e-commerce.}
\label{tab:industrial}
\small
\setlength{\tabcolsep}{2pt}
\renewcommand{\arraystretch}{0.9}
\begin{tabular*}{\columnwidth}{@{\extracolsep{\fill}} l c l c @{}}
\toprule
\multicolumn{2}{c}{\textbf{Panel A: Online retrieval A/B}} &
\multicolumn{2}{c}{\textbf{Panel B: Offline ranking}} \\
\cmidrule(lr){1-2}\cmidrule(lr){3-4}
Target & Gain & Metric & AUC Gain \\
\midrule
GMV    & +0.98\% & Overall CTCVR        & +0.05 pp \\
Orders & +0.91\% & Scenario-A CVR     & +0.05 pp \\
GPM    & +1.16\% & Cold-start CVR & +0.08 pp \\
\bottomrule
\end{tabular*}
\renewcommand{\arraystretch}{1.0}
\end{table}

\subsubsection{Implementation Details}

For all offline experiments, the SID length is set to $3$, and each codebook contains $256$ codes. All methods use the same $768$-dimensional item features as input, and the same downstream TIGER configuration. Unless otherwise specified, the code embedding dimension is set to $32$. We select hyperparameters on the validation set and report the final results on the test set. For AdaSID, we tune a compact set of hyperparameters associated with its three adaptive components, including the upper bound of load-adaptive strengthening $f_{\max}\in\{2.0,3.0\}$, two candidate depth-aware semantic relaxation threshold vectors $\{[0.18,0.24,0.30], [0.14,0.19,0.24]\}$, the adaptive rebalancing start/end steps $\{(10000,200000), (3000,90000)\}$, the collision-weight floor $\lambda_{\mathrm{col}}^{\min}\in\{0.05,0.02\}$, and the collaborative-weight ceiling $\lambda_{\mathrm{cf}}^{\max}\in\{0.25,0.35\}$. The final configuration is chosen on the validation set and then fixed for the reported offline comparisons.

\subsection{Overall Offline Recommendation Performance}

Table~\ref{tab:main_results} compares different SID tokenizers on Toys and Beauty under the same downstream TIGER setting. Since the downstream recommendation backbone, input item features, and evaluation protocol are all fixed, the differences here reflect the quality of the learned SIDs rather than downstream architectural changes. Based on these results, we make the following observations.

\begin{itemize}
    \item AdaSID achieves the best result on all datasets and all metrics. Averaged over the strongest baseline for each metric, the relative improvement is about $6.2\%$ on Toys and $2.9\%$ on Beauty, corresponding to an overall average gain of about $4.5\%$. This pattern suggests that the benefit of AdaSID is not tied to a single dataset or a single evaluation cutoff, but reflects a broadly improved SID quality.

    \item On Toys, AdaSID improves Recall@3 from $0.0195$ to $0.0214$ and NDCG@3 from $0.0164$ to $0.0175$, corresponding to relative gains of about $9.6\%$ and $6.6\%$, respectively. On Beauty, the gain on NDCG@3 is also clear, rising from $0.0155$ to $0.0164$ (about $5.8\%$), whereas the gain on Recall@5 is more modest. This pattern indicates that AdaSID is particularly effective at sharpening fine-grained discriminability in the top-ranked positions rather than merely improving broader retrieval coverage.

    \item Compared with QuaSID, a strong collision-aware baseline, AdaSID improves all four metrics on both datasets, with average relative gains of about $7.4\%$ on Toys and $3.1\%$ on Beauty. AdaSID also surpasses strong code-utilization baselines such as RQ-KMeans and SimRQ on their strongest metrics. This result suggests that the gain does not come merely from adding a collision penalty or from improving generic code usage alone; instead, it comes from regulating overlaps more precisely according to collision load, semantic compatibility, and training stage.
\end{itemize}

\subsection{Industrial Validation}

Table~\ref{tab:industrial} reports industrial validation of AdaSID on Kuaishou e-commerce. In this setting, we first extract multimodal item features from heterogeneous signals such as images, text, keyframes, and audio transcripts using a multimodal foundation model, then train AdaSID based on these features. Specifically, we conduct a 4-day online A/B test on the short-video retrieval model, covering tens of millions of users, and a 4-day offline study on the ranking model. Here, Scenario-A corresponds to a specialized commerce-content slice. Based on these results, we make the following observations.

\begin{itemize}
    \item AdaSID delivers clear business gains in the online A/B test. This result is important because it shows that the benefit of adaptive SID learning is not limited to public offline benchmarks, but can translate into measurable commercial value after deployment in a real recommendation system. In other words, improving semantic discretization on the item side can produce practical gains on both transaction and value-oriented objectives.

    \item The offline ranking results are directionally consistent with the online findings. The observed improvements on both the overall metric and cold-start subsets indicate that the learned SIDs provide useful multimodal item semantics in realistic industrial settings, rather than only working under simplified benchmark conditions. This consistency also strengthens the evidence that the online gains are supported by better item representations.
\end{itemize}

\begin{table}[t]
\centering
\caption{Ablation study of AdaSID on Toys and Beauty. The full model is boldfaced.}
\label{tab:ablation}
\small
\setlength{\tabcolsep}{3.5pt}
\renewcommand{\arraystretch}{0.9}
\begin{tabular}{llcccc}
\toprule
Dataset & Variant & Recall@3 & NDCG@3 & Recall@5 & NDCG@5 \\
\midrule
\multirow{4}{*}{Toys}
& AdaSID        & \textbf{0.0214} & \textbf{0.0175} & \textbf{0.0281} & \textbf{0.0202} \\
& w/o SeAR      & 0.0192 & 0.0153 & 0.0246 & 0.0175 \\
& w/o PAR       & 0.0204 & 0.0169 & 0.0252 & 0.0189 \\
& w/o LAS       & 0.0205 & 0.0161 & 0.0271 & 0.0188 \\
\midrule
\multirow{4}{*}{Beauty}
& AdaSID        & \textbf{0.0205} & \textbf{0.0164} & \textbf{0.0275} & \textbf{0.0190} \\
& w/o SeAR      & 0.0184 & 0.0149 & 0.0263 & 0.0181 \\
& w/o PAR       & 0.0182 & 0.0147 & 0.0236 & 0.0169 \\
& w/o LAS       & 0.0201 & 0.0161 & 0.0269 & 0.0188 \\
\bottomrule
\end{tabular}
\end{table}

\subsection{Codebook Quality Analysis}

Figure~\ref{fig:sid_space_landscape} visualizes the learned SID space by jointly projecting four codebook-side statistics into a single landscape. In this figure, the x-axis corresponds to inverse normalized Top-1 Load, the y-axis corresponds to normalized Minimum Perplexity, point size encodes Mean Perplexity, and point color encodes SID Entropy. A point closer to the upper-right region indicates weaker dominant-code concentration and better weakest-layer utilization, while larger points and points mapped to higher SID entropy indicate stronger overall codebook usage and richer SID-sequence diversity. This joint view is useful because it reveals whether a tokenizer improves the SID space in a balanced manner rather than on only one isolated statistic. Based on Figure~\ref{fig:sid_space_landscape}, we make the following observations.

\begin{itemize}
    \item On Toys, AdaSID lies close to the upper-right frontier and is also among the largest points with high entropy encoding. On Beauty, it again stays near the frontier and remains visually competitive on both size and entropy. This indicates that AdaSID improves weakest-layer utilization and dominant-code concentration without sacrificing the broader diversity of the learned SID space. In other words, its advantage is structural rather than metric-specific.

    \item Compared with strong baselines such as QuaSID, SimRQ, and RQ-KMeans, AdaSID is not better only along one axis while worse on the others. Instead, it remains competitive or favorable simultaneously in position, size, and entropy encoding. This suggests that AdaSID does not obtain better recommendation results by over-optimizing a single codebook statistic; rather, it shifts the overall SID-space profile toward a better joint regime, which is consistent with the ranking improvements in Table~\ref{tab:main_results}.
\end{itemize}

\subsection{Ablation Study}

Table~\ref{tab:ablation} reports the ablation results of AdaSID on Toys and Beauty. SeAR corresponds to the first-stage semantic-adaptive overlap relaxation, whereas LAS and PAR correspond to the second-stage pressure allocation in space and training progress, respectively. We remove one adaptive component at a time while keeping the rest of the tokenizer training pipeline unchanged, in order to test whether the gain of AdaSID comes from the coordinated two-stage design rather than from any single module.

Overall, all three components contribute to the full model, but their roles differ across datasets. Removing LAS causes a consistent yet relatively mild degradation, and this variant remains the strongest ablation on 6 of the 8 reported metrics. This suggests that load-adaptive strengthening is a stable contributor whose main role is to refine collision regulation rather than dominate the overall gain. In contrast, PAR is especially important on Beauty: once it is removed, all four Beauty metrics become the weakest among the ablation variants, indicating that this dataset is particularly sensitive to how optimization emphasis shifts over training. SeAR plays the most critical role on Toys. After removing it, all four Toys metrics drop to the lowest level among the ablation variants, showing that uniform collision suppression alone is insufficient. These results suggest that AdaSID benefits from coordinated adaptive regulation, where semantic-adaptive overlap relaxation prevents over-separation, load-aware strengthening refines local collision handling, and progress-adaptive rebalancing improves recommendation-oriented optimization over training.

\begin{figure}[t]
  \centering
  \includegraphics[width=0.65\columnwidth]{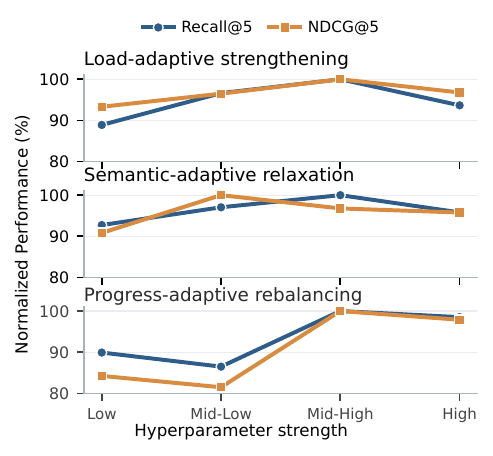}
  \caption{Hyperparameter sensitivity of AdaSID on the Beauty dataset.}
  \label{fig:sensitivity}
\end{figure}

\subsection{Hyperparameter Sensitivity}

Figure~\ref{fig:sensitivity} examines the sensitivity of AdaSID to its three adaptive components on Beauty by varying one component at a time while keeping the others fixed. Overall, AdaSID does not improve monotonically as the adaptive strength increases. The best performance is generally achieved in a moderate-to-strong regime rather than at the weakest or most aggressive setting, indicating that effective SID learning requires calibrated adaptive control. Among the three components, progress-adaptive rebalancing produces the largest performance variation, suggesting that how the optimization emphasis shifts over training is particularly important for downstream recommendation. In contrast, load-adaptive strengthening and semantics-adaptive relaxation show smoother trends, which indicates that their contributions are more stable over a wider range of settings.

\section{Conclusion and Future Work}

In this paper, we proposed AdaSID, an adaptive SID learning framework for recommendation. By treating overlap regulation as a two-stage adaptive process—first relaxing repulsion under semantic compatibility and then regulating the remaining overlaps across space and training time—AdaSID learns SIDs that better balance discrete-space discriminability and downstream recommendation alignment. Extensive experiments on public benchmarks show consistent improvements in both recommendation performance and codebook quality, and industrial validation on Kuaishou e-commerce further confirms its practical value in real-world recommendation scenarios. In the future, it would be interesting to extend adaptive SID learning to joint user--item discrete modeling, tighter end-to-end coupling with generative recommenders, and more scalable settings with longer SIDs, larger codebooks, and richer multimodal inputs.

\bibliographystyle{ACM-Reference-Format}
\bibliography{sample-base}

@inproceedings{10.1145/3640457.3688190,
author = {Singh, Anima and Vu, Trung and Mehta, Nikhil and Keshavan, Raghunandan and Sathiamoorthy, Maheswaran and Zheng, Yilin and Hong, Lichan and Heldt, Lukasz and Wei, Li and Tandon, Devansh and Chi, Ed and Yi, Xinyang},
title = {Better Generalization with Semantic IDs: A Case Study in Ranking for Recommendations},
year = {2024},
isbn = {9798400705052},
publisher = {Association for Computing Machinery},
address = {New York, NY, USA},
url = {https://doi.org/10.1145/3640457.3688190},
doi = {10.1145/3640457.3688190},
pages = {1039–1044},
numpages = {6},
location = {Bari, Italy},
series = {RecSys '24}
}

@inproceedings{10.1145/3705328.3748123,
author = {Zheng, Carolina and Huang, Minhui and Pedchenko, Dmitrii and Rangadurai, Kaushik and Wang, Siyu and Xia, Fan and Nahum, Gaby and Lei, Jie and Yang, Yang and Liu, Tao and Luo, Zutian and Wei, Xiaohan and Ramasamy, Dinesh and Yang, Jiyan and Han, Yiping and Yang, Lin and Xu, Hangjun and Jin, Rong and Yang, Shuang},
title = {Enhancing Embedding Representation Stability in Recommendation Systems with Semantic ID},
year = {2025},
isbn = {9798400713644},
publisher = {Association for Computing Machinery},
address = {New York, NY, USA},
url = {https://doi.org/10.1145/3705328.3748123},
doi = {10.1145/3705328.3748123},
booktitle = {Proceedings of the Nineteenth ACM Conference on Recommender Systems},
pages = {954–957},
numpages = {4},
location = {
},
series = {RecSys '25}
}

@inproceedings{10.1145/3746252.3761439,
author = {Agarwal, Prabhat and Badrinath, Anirudhan and Bhasin, Laksh and Yang, Jaewon and Xu, Jiajing and Rosenberg, Charles},
title = {Autoregressive Generative Retrieval for Industrial-Scale Recommendations at Pinterest},
year = {2025},
isbn = {9798400720406},
publisher = {Association for Computing Machinery},
address = {New York, NY, USA},
url = {https://doi.org/10.1145/3746252.3761439},
doi = {10.1145/3746252.3761439},
booktitle = {Proceedings of the 34th ACM International Conference on Information and Knowledge Management},
pages = {6861–6862},
numpages = {2},
location = {Seoul, Republic of Korea},
series = {CIKM '25}
}

@inproceedings{10.1145/3746252.3761502,
author = {Luo, Xinchen and Cao, Jiangxia and Sun, Tianyu and Yu, Jinkai and Huang, Rui and Yuan, Wei and Lin, Hezheng and Zheng, Yichen and Wang, Shiyao and Hu, Qigen and Qiu, Changqing and Zhang, Jiaqi and Zhang, Xu and Yan, Zhiheng and Zhang, Jingming and Zhang, Simin and Wen, Mingxing and Liu, Zhaojie and Zhou, Guorui},
title = {QARM: Quantitative Alignment Multi-Modal Recommendation at Kuaishou},
year = {2025},
isbn = {9798400720406},
publisher = {Association for Computing Machinery},
address = {New York, NY, USA},
url = {https://doi.org/10.1145/3746252.3761502},
doi = {10.1145/3746252.3761502},
booktitle = {Proceedings of the 34th ACM International Conference on Information and Knowledge Management},
pages = {5915–5922},
numpages = {8},
location = {Seoul, Republic of Korea},
series = {CIKM '25}
}

@inproceedings{10.1145/3746252.3761529,
author = {Ye, Wencai and Sun, Mingjie and Shi, Shaoyun and Wang, Peng and Wu, Wenjin and Jiang, Peng},
title = {DAS: Dual-Aligned Semantic IDs Empowered Industrial Recommender System},
year = {2025},
isbn = {9798400720406},
publisher = {Association for Computing Machinery},
address = {New York, NY, USA},
url = {https://doi.org/10.1145/3746252.3761529},
doi = {10.1145/3746252.3761529},
pages = {6217–6224},
numpages = {8},
location = {Seoul, Republic of Korea},
series = {CIKM '25}
}

@inproceedings{10.1145/3746027.3754937,
author = {Huang, Siyuan and Jin, Jiahui and Lin, Xin and Sun, Xigang and Ban, Yukun},
title = {IM-POI: Bridging ID and Multi-modal Gaps in Next POI Recommendation},
year = {2025},
isbn = {9798400720352},
publisher = {Association for Computing Machinery},
address = {New York, NY, USA},
url = {https://doi.org/10.1145/3746027.3754937},
doi = {10.1145/3746027.3754937},
booktitle = {Proceedings of the 33rd ACM International Conference on Multimedia},
pages = {5979–5987},
numpages = {9},
location = {Dublin, Ireland},
series = {MM '25}
}

@inproceedings{10.1145/3664647.3680574,
author = {Lv, Zheqi and He, Shaoxuan and Zhan, Tianyu and Zhang, Shengyu and Zhang, Wenqiao and Chen, Jingyuan and Zhao, Zhou and Wu, Fei},
title = {Semantic Codebook Learning for Dynamic Recommendation Models},
year = {2024},
isbn = {9798400706868},
publisher = {Association for Computing Machinery},
address = {New York, NY, USA},
url = {https://doi.org/10.1145/3664647.3680574},
doi = {10.1145/3664647.3680574},
booktitle = {Proceedings of the 32nd ACM International Conference on Multimedia},
pages = {9611–9620},
numpages = {10},
location = {Melbourne VIC, Australia},
series = {MM '24}
}

@article{rajput2023recommender,
  title={Recommender systems with generative retrieval},
  author={Rajput, Shashank and Mehta, Nikhil and Singh, Anima and Hulikal Keshavan, Raghunandan and Vu, Trung and Heldt, Lukasz and Hong, Lichan and Tay, Yi and Tran, Vinh and Samost, Jonah and others},
  journal={Advances in Neural Information Processing Systems},
  volume={36},
  pages={10299--10315},
  year={2023}
}

@inproceedings{10.5555/3692070.3692964,
author = {Jin, Bowen and Zeng, Hansi and Wang, Guoyin and Chen, Xiusi and Wei, Tianxin and Li, Ruirui and Wang, Zhengyang and Li, Zheng and Li, Yang and Lu, Hanqing and Wang, Suhang and Han, Jiawei and Tang, Xianfeng},
title = {Language models as semantic indexers},
year = {2024},
publisher = {JMLR.org},
booktitle = {Proceedings of the 41st International Conference on Machine Learning},
articleno = {894},
numpages = {16},
location = {Vienna, Austria},
series = {ICML'24}
}

@article{li2025survey,
  title={A survey of generative recommendation from a tri-decoupled perspective: Tokenization, architecture, and optimization},
  author={Li, Xiaopeng and Chen, Bo and She, Junda and Cao, Shiteng and Wang, You and Jia, Qinlin and He, Haiying and Zhou, Zheli and Liu, Zhao and Liu, Ji and others},
  year={2025},
  publisher={Preprints}
}

@inproceedings{10.1145/3627673.3679569,
author = {Wang, Wenjie and Bao, Honghui and Lin, Xinyu and Zhang, Jizhi and Li, Yongqi and Feng, Fuli and Ng, See-Kiong and Chua, Tat-Seng},
title = {Learnable Item Tokenization for Generative Recommendation},
year = {2024},
isbn = {9798400704369},
publisher = {Association for Computing Machinery},
address = {New York, NY, USA},
url = {https://doi.org/10.1145/3627673.3679569},
doi = {10.1145/3627673.3679569},
booktitle = {Proceedings of the 33rd ACM International Conference on Information and Knowledge Management},
pages = {2400–2409},
numpages = {10},
location = {Boise, ID, USA},
series = {CIKM '24}
}

@inproceedings{10.1145/3726302.3729989,
author = {Liu, Enze and Zheng, Bowen and Ling, Cheng and Hu, Lantao and Li, Han and Zhao, Wayne Xin},
title = {Generative Recommender with End-to-End Learnable Item Tokenization},
year = {2025},
isbn = {9798400715921},
publisher = {Association for Computing Machinery},
address = {New York, NY, USA},
url = {https://doi.org/10.1145/3726302.3729989},
doi = {10.1145/3726302.3729989},
booktitle = {Proceedings of the 48th International ACM SIGIR Conference on Research and Development in Information Retrieval},
pages = {729–739},
numpages = {11},
location = {Padua, Italy},
series = {SIGIR '25}
}

@article{hu2026stop,
  title={Stop Treating Collisions Equally: Qualification-Aware Semantic ID Learning for Recommendation at Industrial Scale},
  author={Hu, Zheng and Chen, Yuxin and Pan, Yongsen and Yuan, Xu and Yin, Yuting and Wang, Daoyuan and Xia, Boyang and Luo, Zefei and Wang, Hongyang and Ni, Songhao and others},
  journal={arXiv preprint arXiv:2603.00632},
  year={2026}
}

@article{van2017neural,
  title={Neural discrete representation learning},
  author={Van Den Oord, Aaron and Vinyals, Oriol and others},
  journal={Advances in neural information processing systems},
  volume={30},
  year={2017}
}

@inproceedings{DBLP:conf/cvpr/LeeKKCH22,
  author       = {Doyup Lee and
                  Chiheon Kim and
                  Saehoon Kim and
                  Minsu Cho and
                  Wook{-}Shin Han},
  title        = {Autoregressive Image Generation using Residual Quantization},
  booktitle    = {{IEEE/CVF} Conference on Computer Vision and Pattern Recognition,
                  {CVPR} 2022, New Orleans, LA, USA, June 18-24, 2022},
  pages        = {11513--11522},
  publisher    = {{IEEE}},
  year         = {2022},
  url          = {https://doi.org/10.1109/CVPR52688.2022.01123},
  doi          = {10.1109/CVPR52688.2022.01123},
}

@inproceedings{Mentzer2024FiniteSQ,
  title={Finite Scalar Quantization: VQ-VAE Made Simple},
  author={Fabian Mentzer and David Minnen and Eirikur Agustsson and Michael Tschannen},
  booktitle={The Twelfth International Conference on Learning Representations (ICLR)},
  year={2024},
  url={https://arxiv.org/abs/2309.15505}
}

@article{DBLP:journals/corr/abs-2508-04618,
  author       = {Dengzhao Fang and
                  Jingtong Gao and
                  Chengcheng Zhu and
                  Yu Li and
                  Xiangyu Zhao and
                  Yi Chang},
  title        = {HiD-VAE: Interpretable Generative Recommendation via Hierarchical
                  and Disentangled Semantic IDs},
  journal      = {CoRR},
  volume       = {abs/2508.04618},
  year         = {2025},
  url          = {https://doi.org/10.48550/arXiv.2508.04618},
  doi          = {10.48550/ARXIV.2508.04618},
  eprinttype    = {arXiv},
  eprint       = {2508.04618},
}

@inproceedings{10.1145/3637528.3671473,
author = {Liu, Qijiong and Zhu, Jieming and Yang, Yanting and Dai, Quanyu and Du, Zhaocheng and Wu, Xiao-Ming and Zhao, Zhou and Zhang, Rui and Dong, Zhenhua},
title = {Multimodal Pretraining, Adaptation, and Generation for Recommendation: A Survey},
year = {2024},
isbn = {9798400704901},
publisher = {Association for Computing Machinery},
address = {New York, NY, USA},
url = {https://doi.org/10.1145/3637528.3671473},
doi = {10.1145/3637528.3671473},
booktitle = {Proceedings of the 30th ACM SIGKDD Conference on Knowledge Discovery and Data Mining},
pages = {6566–6576},
numpages = {11},
location = {Barcelona, Spain},
series = {KDD '24}
}

@inproceedings{10.1145/3746027.3755540,
author = {Yang, Wei and Zhong, Rui and Chen, Yiqun and Li, Shixuan and Ping, Heng and Lu, Chi and Jiang, Peng},
title = {FITMM: Adaptive Frequency-Aware Multimodal Recommendation via Information-Theoretic Representation Learning},
year = {2025},
isbn = {9798400720352},
publisher = {Association for Computing Machinery},
address = {New York, NY, USA},
url = {https://doi.org/10.1145/3746027.3755540},
doi = {10.1145/3746027.3755540},
booktitle = {Proceedings of the 33rd ACM International Conference on Multimedia},
pages = {6193–6202},
numpages = {10},
location = {Dublin, Ireland},
series = {MM '25}
}

@inproceedings{10.1145/3746027.3755781,
author = {Xu, Jinfeng and Chen, Zheyu and Yang, Shuo and Li, Jinze and Ngai, Edith C. H.},
title = {The Best is Yet to Come: Graph Convolution in the Testing Phase for Multimodal Recommendation},
year = {2025},
isbn = {9798400720352},
publisher = {Association for Computing Machinery},
address = {New York, NY, USA},
url = {https://doi.org/10.1145/3746027.3755781},
doi = {10.1145/3746027.3755781},
booktitle = {Proceedings of the 33rd ACM International Conference on Multimedia},
pages = {6325–6334},
numpages = {10},
location = {Dublin, Ireland},
series = {MM '25}
}

@inproceedings{10.1145/3637528.3672041,
author = {Cheng, Zhangtao and Zhang, Jienan and Xu, Xovee and Trajcevski, Goce and Zhong, Ting and Zhou, Fan},
title = {Retrieval-Augmented Hypergraph for Multimodal Social Media Popularity Prediction},
year = {2024},
isbn = {9798400704901},
publisher = {Association for Computing Machinery},
address = {New York, NY, USA},
url = {https://doi.org/10.1145/3637528.3672041},
doi = {10.1145/3637528.3672041},
booktitle = {Proceedings of the 30th ACM SIGKDD Conference on Knowledge Discovery and Data Mining},
pages = {445–455},
numpages = {11},
location = {Barcelona, Spain},
series = {KDD '24}
}

@inproceedings{10.5555/3015812.3015834,
author = {He, Ruining and McAuley, Julian},
title = {VBPR: visual Bayesian Personalized Ranking from implicit feedback},
year = {2016},
publisher = {AAAI Press},
booktitle = {Proceedings of the Thirtieth AAAI Conference on Artificial Intelligence},
pages = {144–150},
numpages = {7},
location = {Phoenix, Arizona},
series = {AAAI'16}
}

@inproceedings{10.1145/3343031.3351034,
author = {Wei, Yinwei and Wang, Xiang and Nie, Liqiang and He, Xiangnan and Hong, Richang and Chua, Tat-Seng},
title = {MMGCN: Multi-modal Graph Convolution Network for Personalized Recommendation of Micro-video},
year = {2019},
isbn = {9781450368896},
publisher = {Association for Computing Machinery},
address = {New York, NY, USA},
url = {https://doi.org/10.1145/3343031.3351034},
doi = {10.1145/3343031.3351034},
booktitle = {Proceedings of the 27th ACM International Conference on Multimedia},
pages = {1437–1445},
numpages = {9},
location = {Nice, France},
series = {MM '19}
}

@inproceedings{10.1145/3640457.3688098,
author = {Chen, Gaode and Sun, Ruina and Jiang, Yuezihan and Cao, Jiangxia and Zhang, Qi and Lin, Jingjian and Li, Han and Gai, Kun and Zhang, Xinghua},
title = {A Multi-modal Modeling Framework for Cold-start Short-video Recommendation},
year = {2024},
isbn = {9798400705052},
publisher = {Association for Computing Machinery},
address = {New York, NY, USA},
url = {https://doi.org/10.1145/3640457.3688098},
doi = {10.1145/3640457.3688098},
booktitle = {Proceedings of the 18th ACM Conference on Recommender Systems},
pages = {391–400},
numpages = {10},
location = {Bari, Italy},
series = {RecSys '24}
}

@inproceedings{10.1145/3474085.3475259,
author = {Zhang, Jinghao and Zhu, Yanqiao and Liu, Qiang and Wu, Shu and Wang, Shuhui and Wang, Liang},
title = {Mining Latent Structures for Multimedia Recommendation},
year = {2021},
isbn = {9781450386517},
publisher = {Association for Computing Machinery},
address = {New York, NY, USA},
url = {https://doi.org/10.1145/3474085.3475259},
doi = {10.1145/3474085.3475259},
booktitle = {Proceedings of the 29th ACM International Conference on Multimedia},
pages = {3872–3880},
numpages = {9},
location = {Virtual Event, China},
series = {MM '21}
}

@inproceedings{10.1145/3512527.3531378,
author = {Liu, Zhuang and Ma, Yunpu and Schubert, Matthias and Ouyang, Yuanxin and Xiong, Zhang},
title = {Multi-Modal Contrastive Pre-training for Recommendation},
year = {2022},
isbn = {9781450392389},
publisher = {Association for Computing Machinery},
address = {New York, NY, USA},
url = {https://doi.org/10.1145/3512527.3531378},
doi = {10.1145/3512527.3531378},
booktitle = {Proceedings of the 2022 International Conference on Multimedia Retrieval},
pages = {99–108},
numpages = {10},
location = {Newark, NJ, USA},
series = {ICMR '22}
}

@article{10.1145/3682075,
author = {Zhang, Lingzi and Zhou, Xin and Zeng, Zhiwei and Shen, Zhiqi},
title = {Multimodal Pre-training for Sequential Recommendation via Contrastive Learning},
year = {2024},
issue_date = {March 2025},
publisher = {Association for Computing Machinery},
address = {New York, NY, USA},
volume = {3},
number = {1},
url = {https://doi.org/10.1145/3682075},
doi = {10.1145/3682075},
journal = {ACM Trans. Recomm. Syst.},
month = oct,
articleno = {9},
numpages = {23}
}

@inproceedings{10.1145/3543507.3583251,
author = {Zhou, Xin and Zhou, Hongyu and Liu, Yong and Zeng, Zhiwei and Miao, Chunyan and Wang, Pengwei and You, Yuan and Jiang, Feijun},
title = {Bootstrap Latent Representations for Multi-modal Recommendation},
year = {2023},
isbn = {9781450394161},
publisher = {Association for Computing Machinery},
address = {New York, NY, USA},
url = {https://doi.org/10.1145/3543507.3583251},
doi = {10.1145/3543507.3583251},
booktitle = {Proceedings of the ACM Web Conference 2023},
pages = {845–854},
numpages = {10},
location = {Austin, TX, USA},
series = {WWW '23}
}

@inproceedings{10.1145/3543507.3583434,
author = {Hou, Yupeng and He, Zhankui and McAuley, Julian and Zhao, Wayne Xin},
title = {Learning Vector-Quantized Item Representation for Transferable Sequential Recommenders},
year = {2023},
isbn = {9781450394161},
publisher = {Association for Computing Machinery},
address = {New York, NY, USA},
url = {https://doi.org/10.1145/3543507.3583434},
doi = {10.1145/3543507.3583434},
booktitle = {Proceedings of the ACM Web Conference 2023},
pages = {1162–1171},
numpages = {10},
location = {Austin, TX, USA},
series = {WWW '23}
}

@INPROCEEDINGS{10597986,
  author={Zheng, Bowen and Hou, Yupeng and Lu, Hongyu and Chen, Yu and Zhao, Wayne Xin and Chen, Ming and Wen, Ji-Rong},
  booktitle={2024 IEEE 40th International Conference on Data Engineering (ICDE)}, 
  title={Adapting Large Language Models by Integrating Collaborative Semantics for Recommendation}, 
  year={2024},
  volume={},
  number={},
  pages={1435-1448},
  doi={10.1109/ICDE60146.2024.00118}}

@inproceedings{10.1145/3701716.3715234,
author = {Zhao, Rui and Zhong, Rui and Zheng, Haoran and Yang, Wei and Lu, Chi and Jin, Beihong and Jiang, Peng and Gai, Kun},
title = {Hierarchical Sequence ID Representation of Large Language Models for Large-scale Recommendation Systems},
year = {2025},
isbn = {9798400713316},
publisher = {Association for Computing Machinery},
address = {New York, NY, USA},
url = {https://doi.org/10.1145/3701716.3715234},
doi = {10.1145/3701716.3715234},
booktitle = {Companion Proceedings of the ACM on Web Conference 2025},
pages = {641–650},
numpages = {10},
location = {Sydney NSW, Australia},
series = {WWW '25}
}

@inproceedings{10.1145/3711896.3736979,
author = {Hou, Yupeng and Li, Jiacheng and Shin, Ashley and Jeon, Jinsung and Santhanam, Abhishek and Shao, Wei and Hassani, Kaveh and Yao, Ning and McAuley, Julian},
title = {Generating Long Semantic IDs in Parallel for Recommendation},
year = {2025},
isbn = {9798400714542},
publisher = {Association for Computing Machinery},
address = {New York, NY, USA},
url = {https://doi.org/10.1145/3711896.3736979},
doi = {10.1145/3711896.3736979},
booktitle = {Proceedings of the 31st ACM SIGKDD Conference on Knowledge Discovery and Data Mining V.2},
pages = {956–966},
numpages = {11},
location = {Toronto ON, Canada},
series = {KDD '25}
}

@inproceedings{10.1145/3711896.3737275,
author = {Wu, Binrui and Tang, Shisong and Li, Fan and Han, Bing and Meng, Chang and Xiao, Jingyu and Gao, Jiechao},
title = {Aligning and Balancing ID and Multimodal Representations for Recommendation},
year = {2025},
isbn = {9798400714542},
publisher = {Association for Computing Machinery},
address = {New York, NY, USA},
url = {https://doi.org/10.1145/3711896.3737275},
doi = {10.1145/3711896.3737275},
booktitle = {Proceedings of the 31st ACM SIGKDD Conference on Knowledge Discovery and Data Mining V.2},
pages = {5029–5038},
numpages = {10},
location = {Toronto ON, Canada},
series = {KDD '25}
}

@article{chen2025onesearch,
  title={Onesearch: A preliminary exploration of the unified end-to-end generative framework for e-commerce search},
  author={Chen, Ben and Guo, Xian and Wang, Siyuan and Liang, Zihan and Lv, Yue and Ma, Yufei and Xiao, Xinlong and Xue, Bowen and Zhang, Xuxin and Yang, Ying and others},
  journal={arXiv preprint arXiv:2509.03236},
  year={2025}
}

@article{DBLP:journals/corr/abs-2010-05525,
  author       = {Xiaoyong Yang and
                  Yadong Zhu and
                  Yi Zhang and
                  Xiaobo Wang and
                  Quan Yuan},
  title        = {Large Scale Product Graph Construction for Recommendation in E-commerce},
  journal      = {CoRR},
  volume       = {abs/2010.05525},
  year         = {2020},
  url          = {https://arxiv.org/abs/2010.05525},
  eprinttype    = {arXiv},
  eprint       = {2010.05525}
}

@article{bengio2013estimating,
  title={Estimating or propagating gradients through stochastic neurons for conditional computation},
  author={Bengio, Yoshua and L{\'e}onard, Nicholas and Courville, Aaron},
  journal={arXiv preprint arXiv:1308.3432},
  year={2013}
}

@article{DBLP:journals/corr/abs-1807-03748,
  author       = {A{\"{a}}ron van den Oord and
                  Yazhe Li and
                  Oriol Vinyals},
  title        = {Representation Learning with Contrastive Predictive Coding},
  journal      = {CoRR},
  volume       = {abs/1807.03748},
  year         = {2018},
  url          = {http://arxiv.org/abs/1807.03748},
  eprinttype    = {arXiv},
  eprint       = {1807.03748},
  bibsource    = {dblp computer science bibliography, https://dblp.org}
}

@inproceedings{DBLP:conf/acl/NiACMHCY22,
  author       = {Jianmo Ni and
                  Gustavo Hern{\'{a}}ndez {\'{A}}brego and
                  Noah Constant and
                  Ji Ma and
                  Keith B. Hall and
                  Daniel Cer and
                  Yinfei Yang},
  editor       = {Smaranda Muresan and
                  Preslav Nakov and
                  Aline Villavicencio},
  title        = {Sentence-T5: Scalable Sentence Encoders from Pre-trained Text-to-Text
                  Models},
  booktitle    = {Findings of the Association for Computational Linguistics: {ACL} 2022,
                  Dublin, Ireland, May 22-27, 2022},
  series       = {Findings of {ACL}},
  volume       = {{ACL} 2022},
  pages        = {1864--1874},
  publisher    = {Association for Computational Linguistics},
  year         = {2022},
  url          = {https://doi.org/10.18653/v1/2022.findings-acl.146},
  doi          = {10.18653/V1/2022.FINDINGS-ACL.146},
  bibsource    = {dblp computer science bibliography, https://dblp.org}
}

@inproceedings{DBLP:conf/iclr/YuLKZPQKXBW22,
  author       = {Jiahui Yu and
                  Xin Li and
                  Jing Yu Koh and
                  Han Zhang and
                  Ruoming Pang and
                  James Qin and
                  Alexander Ku and
                  Yuanzhong Xu and
                  Jason Baldridge and
                  Yonghui Wu},
  title        = {Vector-quantized Image Modeling with Improved {VQGAN}},
  booktitle    = {The Tenth International Conference on Learning Representations, {ICLR}
                  2022, Virtual Event, April 25-29, 2022},
  publisher    = {OpenReview.net},
  year         = {2022},
  url          = {https://openreview.net/forum?id=pfNyExj7z2},
}

@inproceedings{DBLP:conf/iclr/FiftyJDILATR25,
  author       = {Christopher Fifty and
                  Ronald Guenther Junkins and
                  Dennis Duan and
                  Aniketh Iyengar and
                  Jerry Weihong Liu and
                  Ehsan Amid and
                  Sebastian Thrun and
                  Christopher R{\'{e}}},
  title        = {Restructuring Vector Quantization with the Rotation Trick},
  booktitle    = {The Thirteenth International Conference on Learning Representations,
                  {ICLR} 2025, Singapore, April 24-28, 2025},
  publisher    = {OpenReview.net},
  year         = {2025},
  url          = {https://openreview.net/forum?id=GMwRl2e9Y1},
}

@article{DBLP:journals/corr/abs-2305-02765,
  author       = {Dongchao Yang and
                  Songxiang Liu and
                  Rongjie Huang and
                  Jinchuan Tian and
                  Chao Weng and
                  Yuexian Zou},
  title        = {HiFi-Codec: Group-residual Vector quantization for High Fidelity Audio
                  Codec},
  journal      = {CoRR},
  volume       = {abs/2305.02765},
  year         = {2023},
  url          = {https://doi.org/10.48550/arXiv.2305.02765},
  doi          = {10.48550/ARXIV.2305.02765},
  eprinttype    = {arXiv},
  eprint       = {2305.02765},
}

@inproceedings{zhu2025addressing,
  title={Addressing representation collapse in vector quantized models with one linear layer},
  author={Zhu, Yongxin and Li, Bocheng and Xin, Yifei and Xia, Zhihua and Xu, Linli},
  booktitle={Proceedings of the IEEE/CVF International Conference on Computer Vision},
  pages={22968--22977},
  year={2025}
}

@inproceedings{DBLP:conf/recsys/Geng0FGZ22,
  author       = {Shijie Geng and
                  Shuchang Liu and
                  Zuohui Fu and
                  Yingqiang Ge and
                  Yongfeng Zhang},
  title        = {Recommendation as Language Processing {(RLP):} {A} Unified Pretrain,
                  Personalized Prompt {\&} Predict Paradigm {(P5)}},
  booktitle    = {RecSys '22: Sixteenth {ACM} Conference on Recommender Systems, Seattle,
                  WA, USA, September 18 - 23, 2022},
  pages        = {299--315},
  publisher    = {{ACM}},
  year         = {2022},
  url          = {https://doi.org/10.1145/3523227.3546767},
  doi          = {10.1145/3523227.3546767}
}

\end{document}